\documentclass[a4paper,twocolumn]{Gaia2004}
\usepackage{epsfig}
\usepackage{graphicx}
\usepackage{amsmath,amsfonts,amssymb,amsxtra} 
\usepackage{natbib}
\usepackage[english]{babel}
\usepackage{times}
\usepackage{url}

\title{A PHOENIX Model Atmosphere Grid for GAIA}
\author[1,2]{I. Brott} 
\author[1]{\& P.H. Hauschildt}
\affil[1]{Hamburger Sternwarte, Gojenbergsweg 112, 21029 Hamburg, Germany}
\affil[2]{INTEGRAL Science Data Centre, Chemin d'Ecogia 16, 1290 Versoix, Switzerland}

\bibpunct{(}{)}{;}{a}{}{,}
\begin{document} 
\maketitle

\begin{abstract}
We present the results of a set of model atmospheres and synthetic
spectra computed with the PHOENIX code. The models cover a range
of effective temperatures (2\,700\,K $\leq T_{\rm{eff}} \leq$ 10\,000\,K), gravities
($-$0.5 $\leq$ $\log(g)$ $\leq$ 5.5) and metalicities ($-$4.0 $\leq
[Z/H] \leq$ $+$0.5). 
In addition, variations of alpha elements are considered for each
metalicity.
The models are computed with a homogeneous set of input data in order
to allow for direct relative comparison between the models. For
example, all models use a mixing length of $l/H_p$ = 2.0. We
provide synthetic spectra with a resolution of 0.2\,nm from the UV to the
infrared for all models. We give a brief overview of the
input physics and show illustrative results. All synthetic spectra are available
via ftp: \url{ftp.hs.uni-hamburg.de/pub/outgoing/phoenix/GAIA} . 
\end{abstract}

\begin{section}{Introduction}
Model atmospheres and synthetic spectra are of fundamental importance 
in the development of filter and resolution systems and also helpful
for the design of the analysis packages for the GAIA mission. 
Here, we present our first results of a grid of model atmospheres and synthetic
spectra computed with the PHOENIX v13 code \cite[]{1999jocaam}. The grid covers the parameter range
of interest to the GAIA community (see Table \ref{tab_gridpar}), $\sim$ 44\,000 models
and spectra in total. It is publicly available via
ftp: \url{ftp.hs.uni-hamburg.de/pub/outgoing/phoenix/GAIA}.

\begin{table}[!bth]
\begin{tabular}{llr}
\hline\hline
parameter & range  & step \\
\hline
$T_{\rm{eff}}$ & 2\,700 $\ldots$ 5\,000\,K & 100\,K \\
         & 5\,000 $\ldots$ 10\,000 K & 200\,K \\
$\log$ g  & $-$0.5 $\ldots$  5.5 & 0.5\\
$[Z/H]$    & $+$0.5 $\ldots$ $-$3.5 & 0.5\\
$[\alpha/\alpha_{\odot}]$ & $-$0.2 $\ldots$ $+$0.8 & 0.2\\
\hline\hline
\end{tabular}
\caption{\label{tab_gridpar}Grid Parameters.}
\end{table}

\begin{subsection}{The PHOENIX code}
\begin{figure}[!htb]
\begin{center}
\psfig{figure=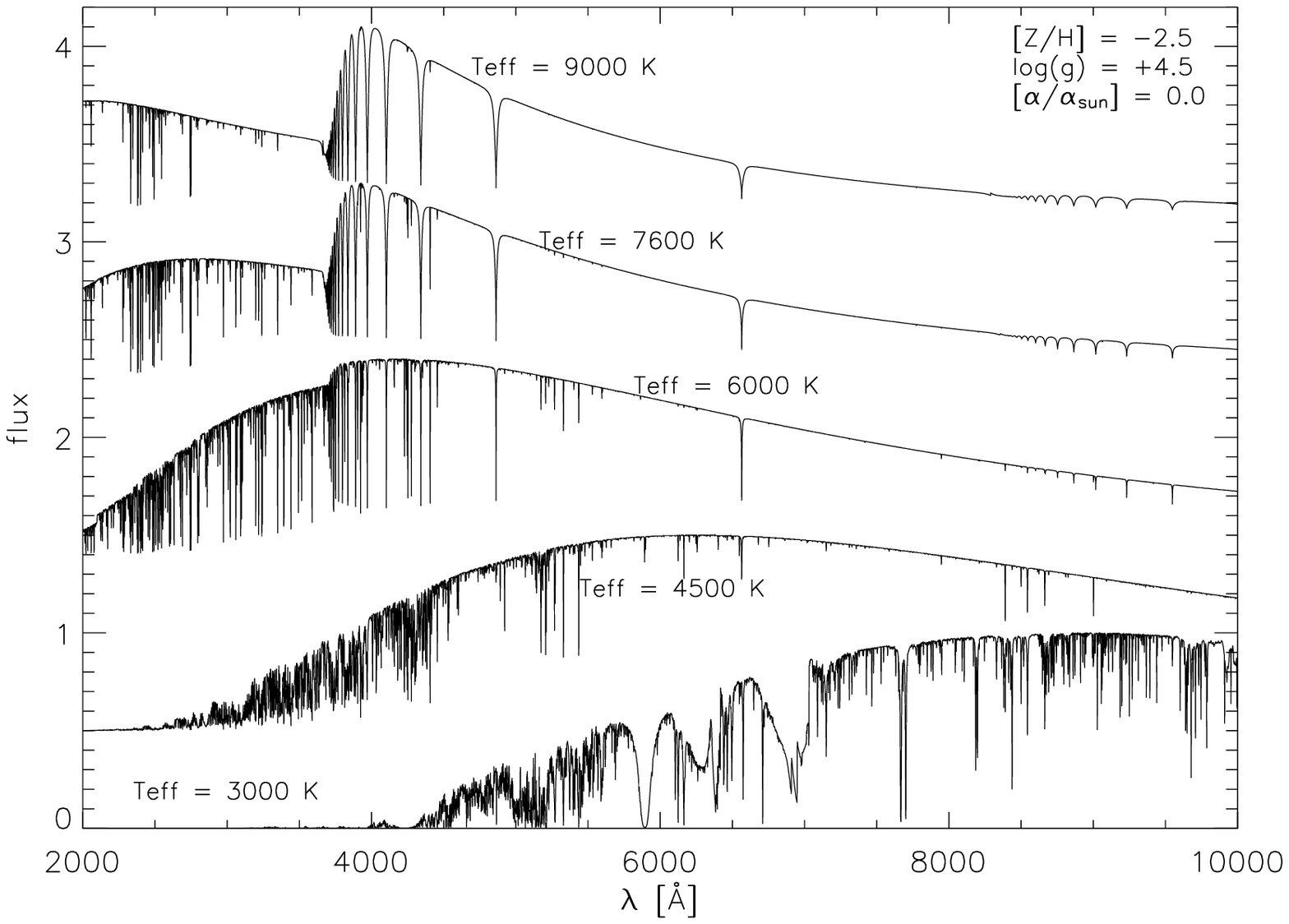,width= \linewidth}
\psfig{figure=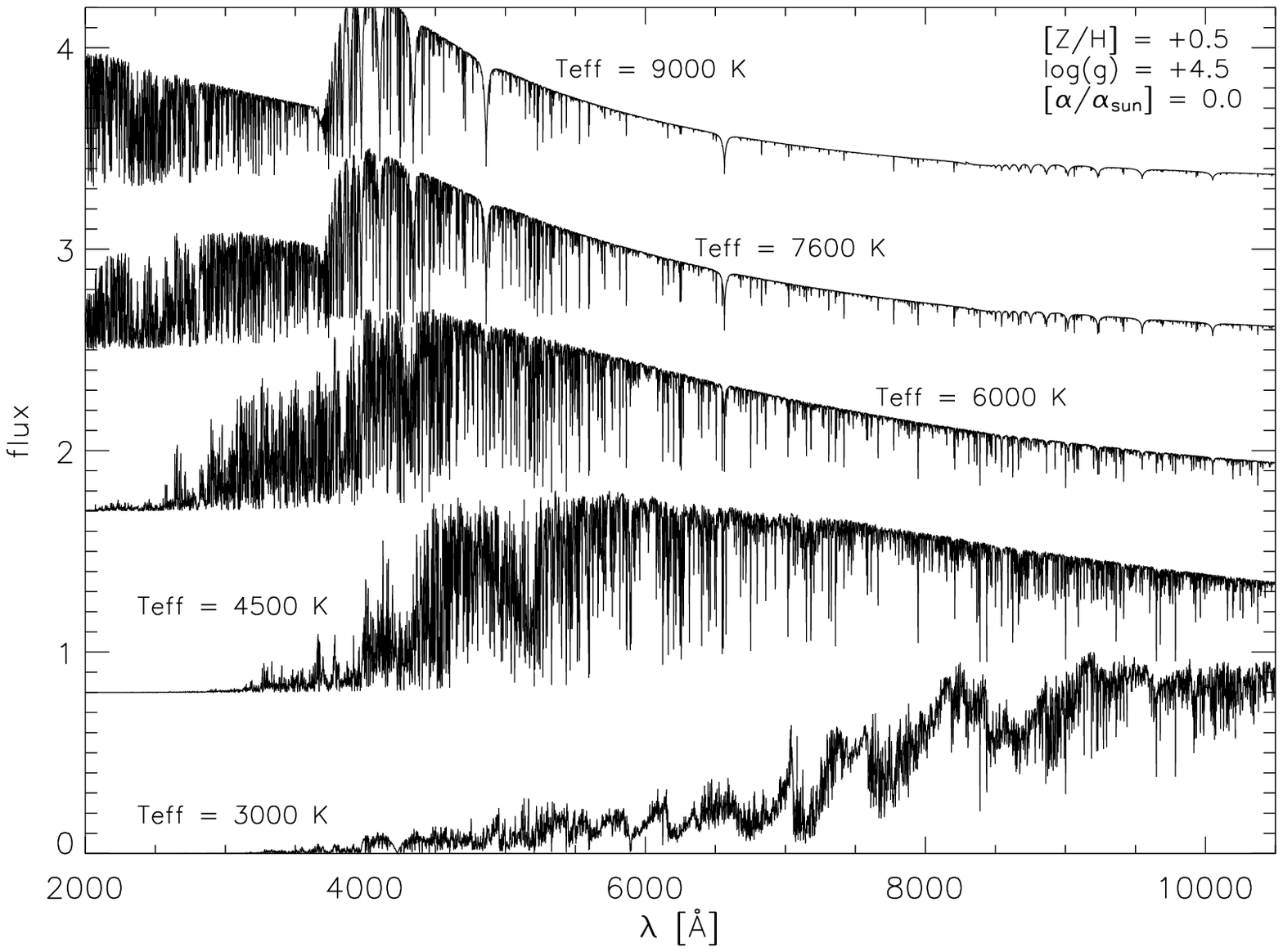,width= \linewidth}
\end{center}
\caption{The effect of temperature variation on a model with
different metalicities and parameters as indicated in the figure. 
At lower temperatures ($<$ 4\,000\,K) metals become important opacity sources
especially in the UV, shifting the flux maximum of the spectrum to longer wavelengths, while 
at higher temperatures ($>$ 7\,000\,K) metals are ionized such that hydrogen is 
dominating the spectrum.}
\label{fig_teff}
\end{figure}

PHOENIX is a multi-purpose stellar atmosphere code
used to compute A-T dwarfs, giants, O-B stars and stellar
winds, exoplanet atmospheres (also irradiated), nova atmospheres, SNe I
\& II and CVs. It was developed by our group to
be a very general NLTE stellar atmosphere code
\cite[]{1992peter,1993peter,1995peter,1995allard,1996Baron,1996Peter,1997Peter,1998Baron,LimDust,2001Peter} . 
The code can handle extremely
large model atoms as well as line blanketing by many millions of
atomic and molecular lines. The radiative transfer in PHOENIX is
solved in spherical geometry and includes the effects of special
relativity (including advection and aberration) in the models. The
PHOENIX code also allows us to include a large number of NLTE and LTE
background lines and to solve the radiative transfer equation for each
of them \textit{without} using simplifying approximations. Therefore,
the line profiles must be resolved in the co-moving (Lagrangian)
frame, which requires many wavelength points (typically 50\,000 in the 
GAIA grid). Details of the numerical methods used in PHOENIX can be 
found in the literature cited above.
In order to take advantage of the enormous computing power and vast
memory sizes of modern parallel computers, we have developed a
parallel version of PHOENIX which is now the default production version
for all our model grids.
\end{subsection}
\end{section} 

\begin{section}{Model Assumptions}
\begin{figure}[!htb]
\begin{center}
\psfig{figure=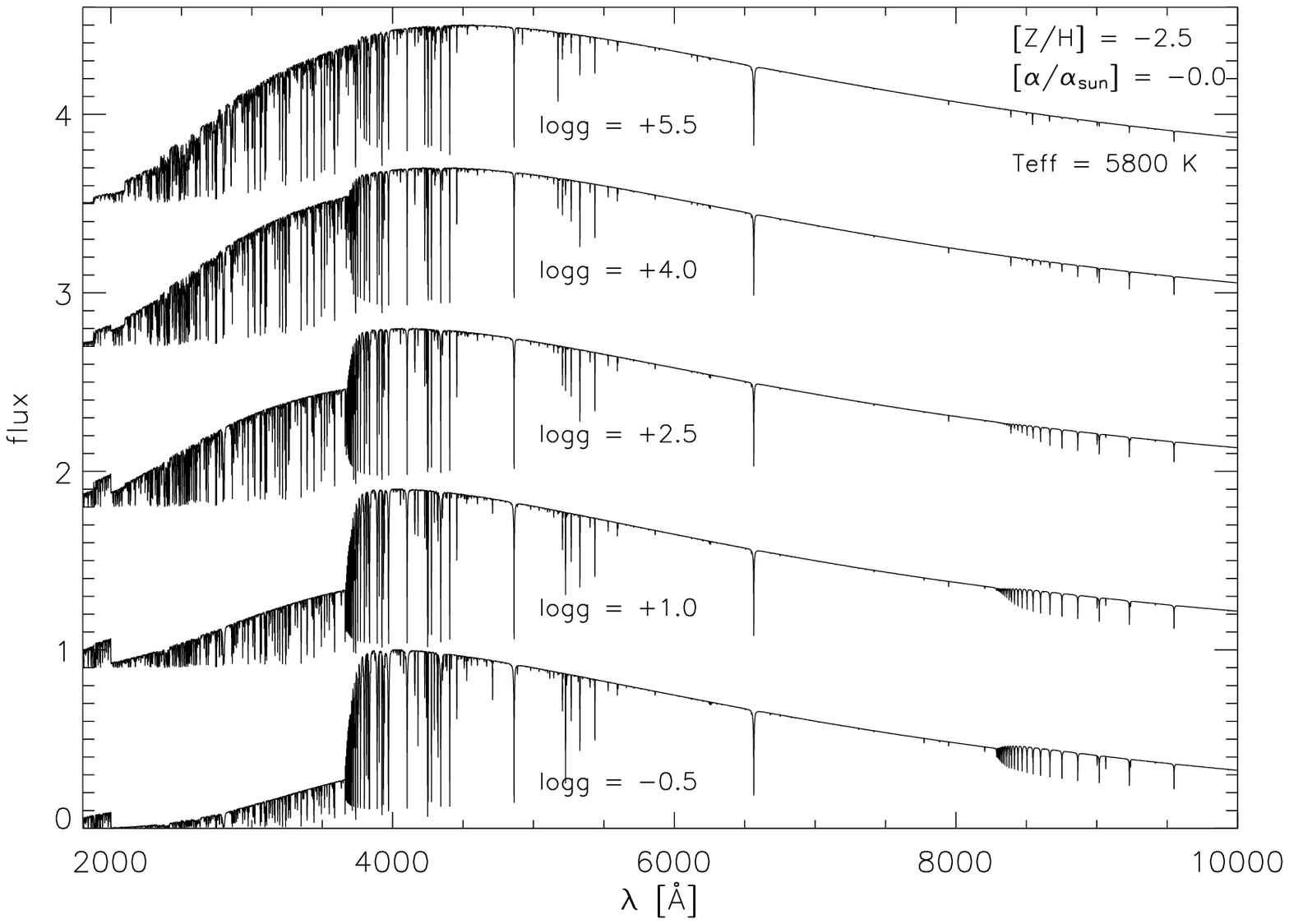,width= \linewidth}
\psfig{figure=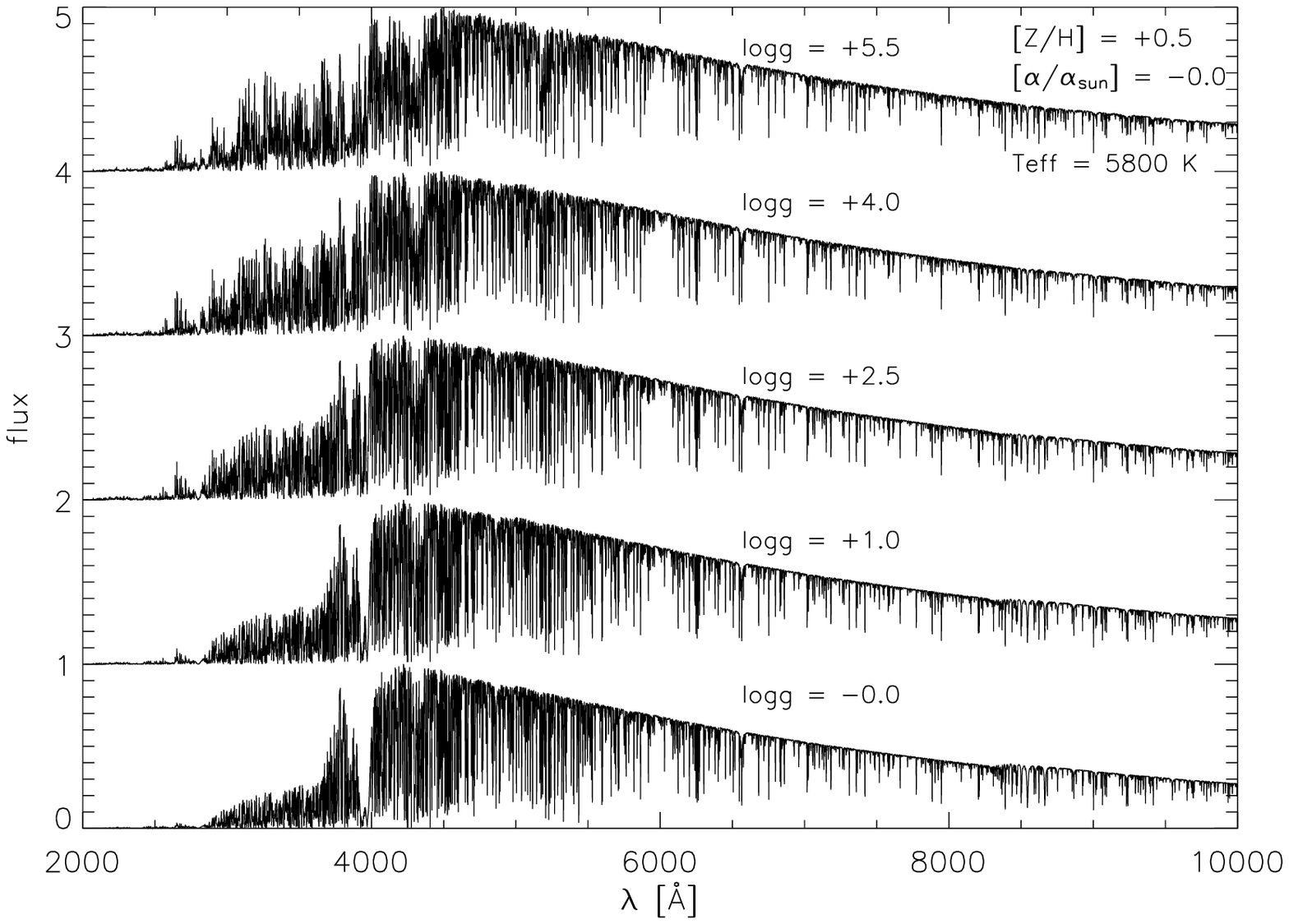,width= \linewidth}
\psfig{figure=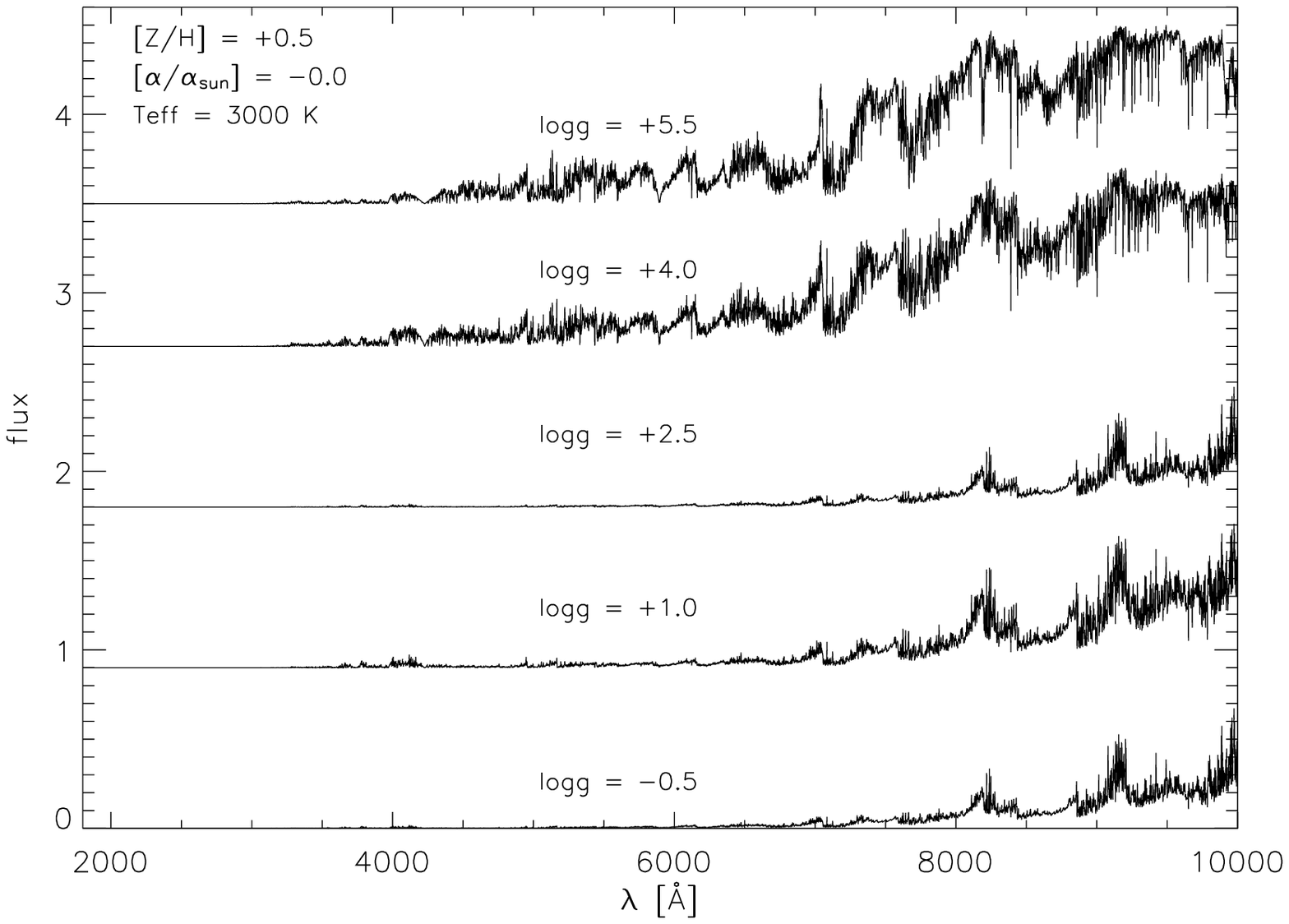,width= \linewidth}
\caption{By varying the gravity, we observe different effects. At high
temperatures (top panel) the Balmer jump gets more pronounced with
lower gravities. At high metalicities it can have an
important effect on molecular lines as shown in the
lower panel for a $[Z/H]$ =$+$0.5, $T_{\rm{eff}}$ = 3\,000\,K model, because
the high pressure influences the EOS, resulting in larger concentration
of molecules. The middle panel shows an intermediate state.}
\label{fig_logg}
\end{center}
\end{figure}

\begin{figure}[!htb]
\centering\psfig{figure=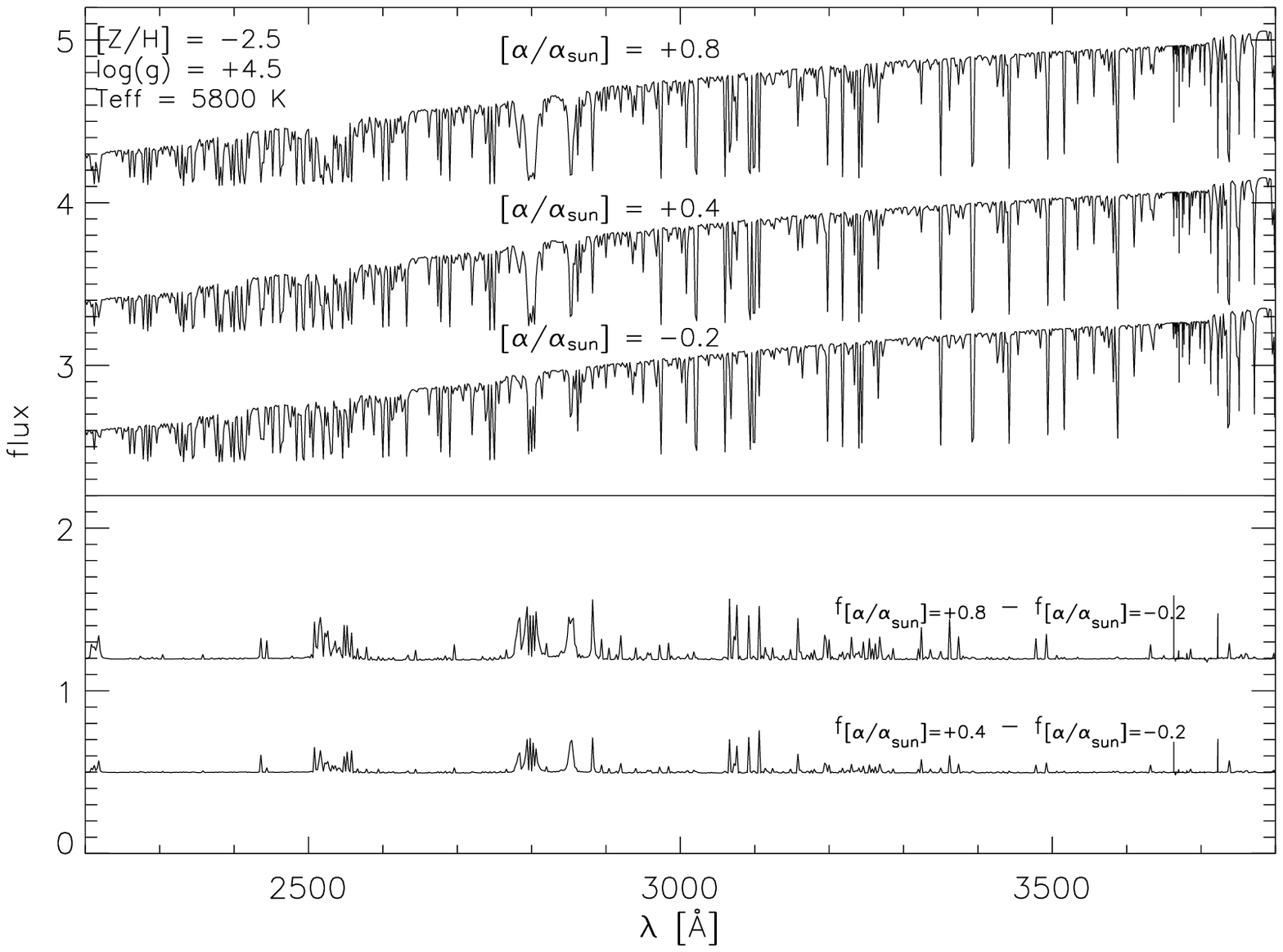,width= \linewidth}
\centering\psfig{figure=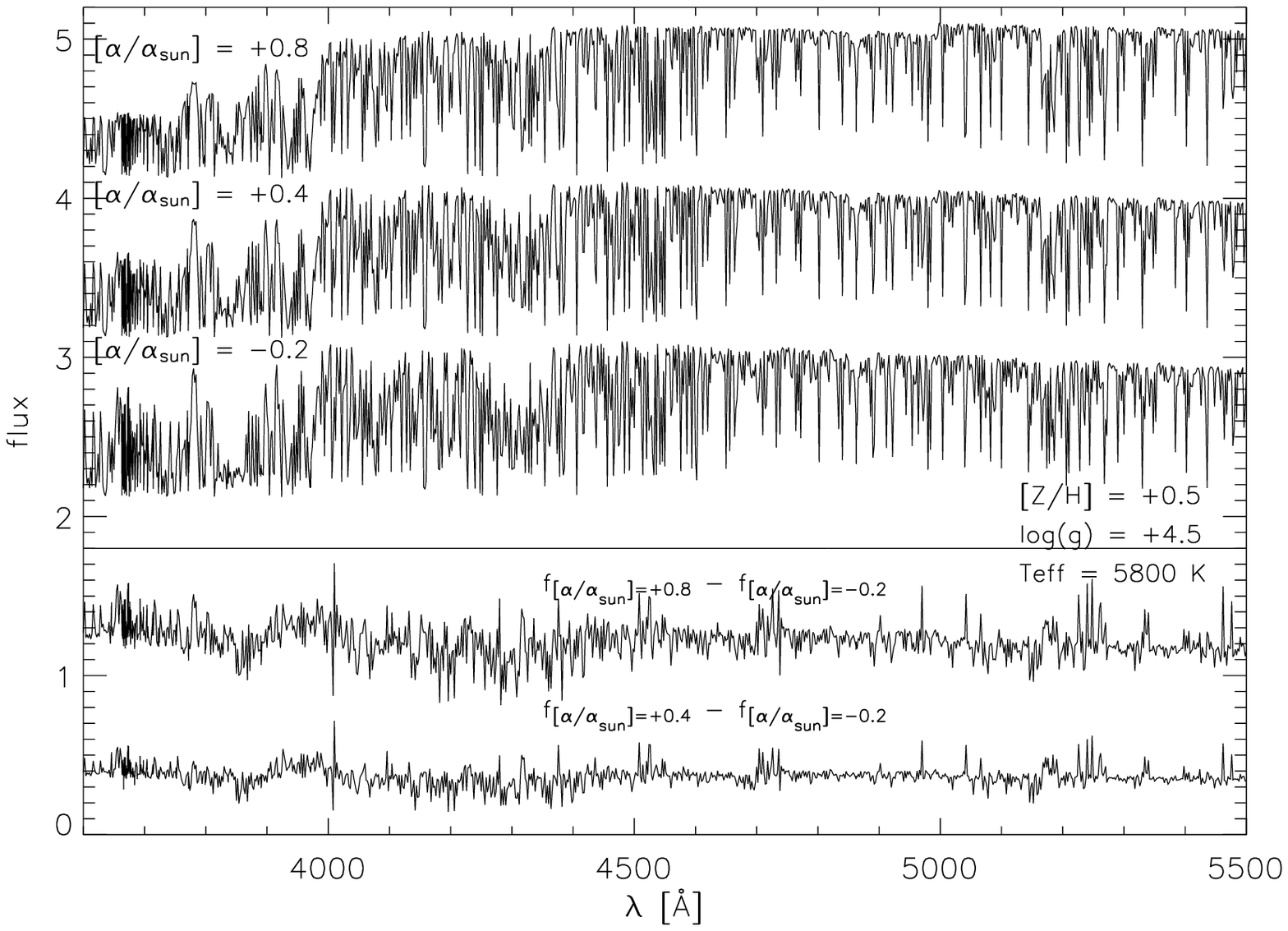,width= \linewidth}
\centering\psfig{figure=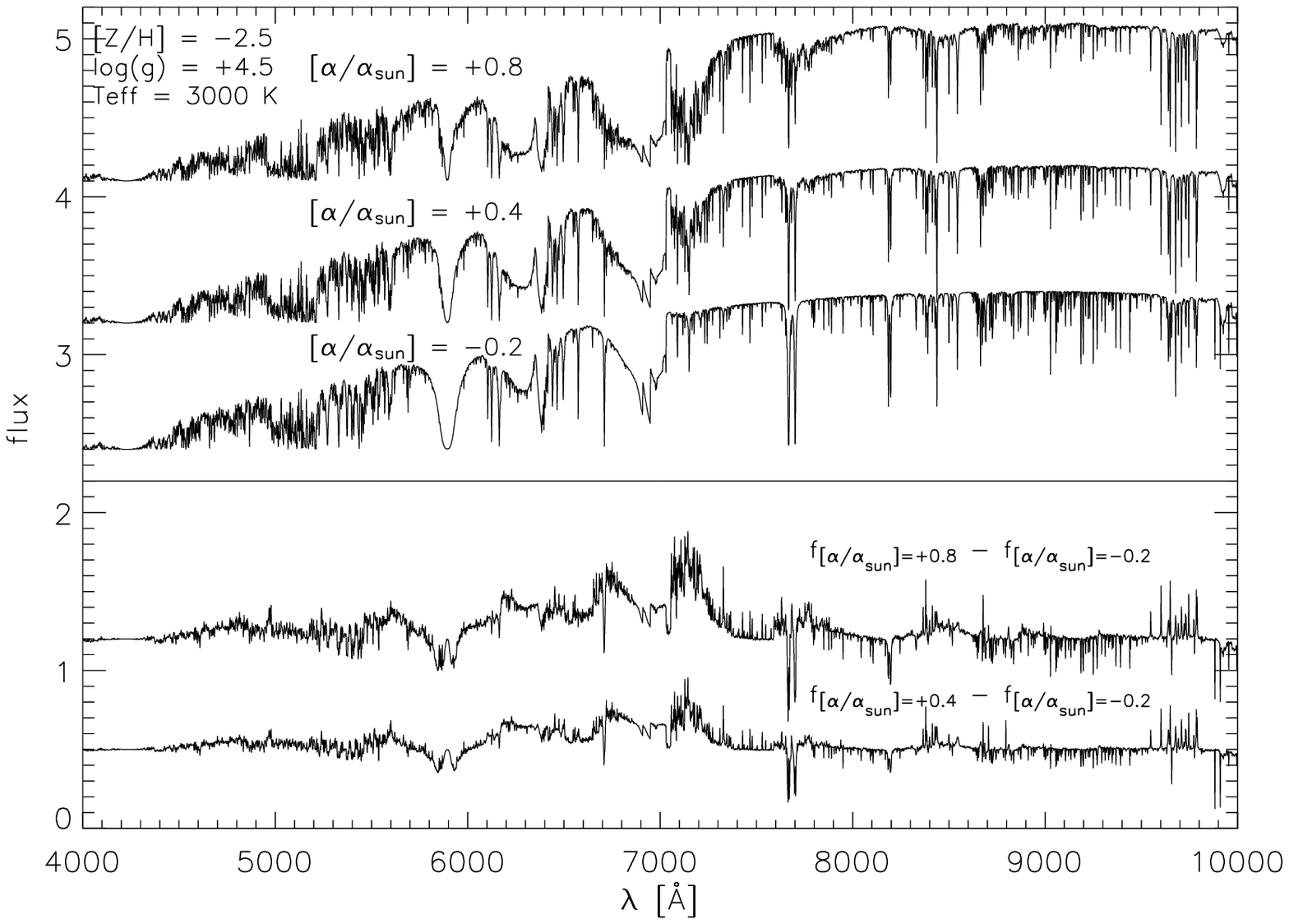,width= \linewidth}
\caption{The variation of the spectrum with changing $\alpha$-element
abundance (compared to the solar value). 
The top panel shows a model with $[Z/H]$ = $-2.5$.
To emphasize the effect of $\alpha$-element abundances we show the difference spectra in the lower part of each panel.
The middle panel shows a model with the same parameters as above but a higher metalicity $[Z/H]$ = $+0.5$, and the lower panel a 3\,000\,K model. This shows that $\alpha$-elements become more important in cool and/or metal rich atmospheres. Note that cool models are influenced over larger wavelengths ranges. }  
\label{fig_alpha1}
\end{figure}

The models of the GAIA grid are calculated for \textbf{1D spherical symmetry} and a fixed stellar mass of 1
$\rm{M}_\odot$ for simplicity. 
Our basic assumptions for the grid presented here are time independence, hydrostatic and thermal
equilibrium, and LTE. Convection is treated with a mixing length of
$l/H_p$ = 2.0. \\
The micro physics setup is similar to the one used in our latest model grids
\cite[]{LimDust}, which give the currently best fits
to the observed spectra of M, L, and T dwarfs.
Water- and TiO-lines  are taken from the AMES calculation from
\cite{partridge97} and \cite{schwenke98}.
Our combined molecular line lists include about 700 million molecular
lines. Typically about 15-300 million of these are selected in a model.  
Atomic and molecular line opacities are selected dynamically for the
relevant LTE background lines from the master line lists and sum the
contribution of every line within a search window to compute the total
line opacity at \textit{arbitrary} wavelength points. 
We do not use pre-computed opacity tables but individual line profiles.
This approach
allows detailed and depth dependent line profiles to be used during the
iterations. This is important i.e. for M dwarf spectra where line
broadening and blanketing are crucial for the model structure and
therefore the computation of the synthetic spectra.  
For details see \cite{2003gsst}.\\
In the present models we have included a constant statistical velocity
field, $\chi = 2 \rm{\,km\,s}^{-1}$, which is treated like a
microturbulence. Lines are chosen if they are stronger than a
threshold, $\Gamma \equiv \xi_l/\kappa_c = 10^{-4}$, where $\xi_l$ and $\kappa_c$ are the extinction coefficient of the line at
its center and the local $b\rm{\,-\,}f$ absorption coefficient, respectively. Details of the line selection process are given in \cite{1999nextgena,1999nextgenb}.\\
The equation of state (EOS) is an enlarged and enhanced version of the
EOS used in \cite{1995allard}. About 1000 species (atoms,
ions, molecules) are included. For cooler models, the formation of
dust particles has to be considered in the EOS.
In our cooler models we have used the assumption, that dust forms and
immediately rains out completely below the photosphere.
For effective temperatures above $2\,700\,$K, dust formation can be 
neglected and therefore this assumption is uncritical for the GAIA grid.
\end{section}

\begin{section}{The GAIA Grid}
In the following, we will show some examples from the GAIA grid.  
Varying one model parameter while keeping the others fixed, we generated a series of plots to demonstrate the behavior of the stellar spectrum. A review about the quality of the synthetic spectra is given in \cite{arunas2004}.

Fig.\ref{fig_teff} shows the effect of temperature variation on a model atmosphere. At lower temperatures metals become important opacity sources especially in the UV, shifting the flux maximum of the spectrum to longer wavelengths. 
At higher temperatures metals are ionized and the spectrum is dominated by hydrogen. 
This effect is independent from the metalicity (upper and lower part of Fig.\ref{fig_teff}).

Increasing the metalicity as shown in Fig.~\ref{fig_zh}, resulting  metal lines dominate the spectrum especially in the UV region.
   
By varying the gravity, we observe different effects (Fig.~\ref{fig_logg}): Hot, metal poor models show a more pronounced Balmer jump with lower gravities (top panel), whereas in metal rich models (middle and lower panel) the gravity can have  important effects on the formation of molecular lines. The high pressure influences the EOS, resulting in a larger concentration of molecules.
     
The GAIA grid also includes different $\alpha$-element abundances for each set of parameters. Some are shown in Fig. \ref{fig_alpha1}. We have varied the abundance of the $\alpha$-elements O, Ne, Mg, Si,
S, Ca, Ti in our models. The effect of $\alpha$-element abundances is small, so each panel shows also the difference spectra as well.
Comparing the middle and lower panel of Fig.\ref{fig_alpha1} one sees, that $\alpha$-element abundances become important at low temperatures and/or high metalicities. 

\begin{figure}[!htb]
\begin{center}
\psfig{figure=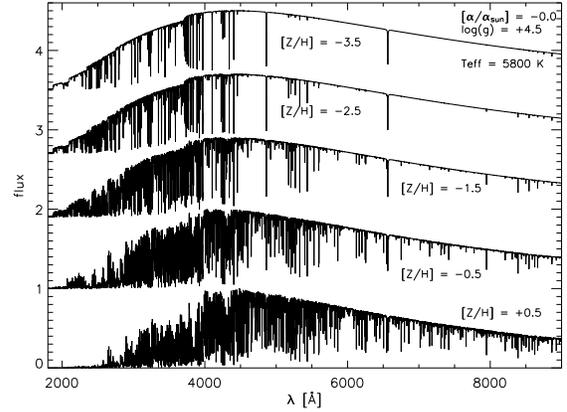,width= \linewidth}
\end{center}
\caption{Models with 
$T_{\rm{eff}}$ = $5800$K and different metalicity. With increasing
metalicity more and more metal lines appear. Especially in the UV
the spectrum is dominated by metal opacity.} 
\label{fig_zh}
\end{figure}
\end{section}

\begin{section}{Conclusions \& Outlook}
The present GAIA grid provides a very detailed set of synthetic spectra to
support the photometric working group.  We plan a full extension of the
grid to the temperature range between 10\,000\,K and 50\,000\,K. These calculations
will be done in full NLTE.  It is also possible to extend the present synthetic
spectra to higher resolution upon request to the authors.
\end{section}

\begin{section}*{Acknowledgments}
PHH was supported in part by the P\^ole Scientifique de Mod\'elisation
Num\'erique at ENS-Lyon. Some of the calculations presented here were performed
at the H\"ochstleistungs Rechenzentrum Nord (HLRN), at the National Energy
Research Supercomputer Center (NERSC), supported by the U.S. DOE, and at the
San Diego Supercomputer Center (SDSC), supported by the NSF.  We thank all
these institutions for a generous allocation of computer time.
\end{section}

\bibliography{literatur}

\end{document}